\documentclass[conference]{IEEEtran}

\usepackage{cite}
\usepackage{amsmath,amssymb,amsfonts}
\usepackage{algorithmic}
\usepackage{graphicx}
\usepackage{textcomp}
\usepackage{xcolor}
\usepackage{multirow}
\usepackage{array}
\usepackage{booktabs}
\usepackage{amsthm}

\usepackage{pifont}
\def\BibTeX{{\rm B\kern-.05em{\sc i\kern-.025em b}\kern-.08em
    T\kern-.1667em\lower.7ex\hbox{E}\kern-.125emX}}
\newcommand{\projectname}{PointZip}
\begin{document}

\title{
\projectname{}: Statistically Lossless Compression for Climate Data with Pointwise Error Bounds
}

\author{
\IEEEauthorblockN{
Jiayong Li, Pratyai Mazumder, Yuejiang Yu, Langwen Huang, Alexandru Calotoiu, Torsten Hoefler
}
\IEEEauthorblockA{
ETH Zurich, Zurich, Switzerland \\
\{jiayong.li, pratyai.mazumder, yuejiang.yu, langwen.huang, alexandru.calotoiu, torsten.hoefler\}@inf.ethz.ch
}
}

\maketitle
\thispagestyle{plain}
\pagestyle{plain}

\begin{abstract}
High-resolution climate simulations generate data volumes that 
increasingly strain storage, I/O, and data dissemination.
Error-bounded lossy compressors can reduce these costs, 
but their tolerances are usually user-chosen 
and difficult to standardize across heterogeneous climate datasets. 
We address this limitation with statistically lossless compression, 
where the pointwise compression error is bounded by the pointwise data uncertainty. 
We propose \projectname{}, a two-stage compressor that combines a rate-distortion-efficient base layer 
with error-bound-aware residual coding to enforce pointwise guarantees. 
On outputs from the widely used climate modeling framework ICON, 
\projectname{} makes 1km simulations practical by reducing storage bandwidth requirements by $72\times$.
On the ERA5 climate dataset, \projectname{} reaches $69.3\times$ on the most representative variables, reducing the commonly used subset to 4~TB. 
In downstream evaluations, \projectname{} preserves data spectra,
extreme events, and air parcel trajectories,
while increasing neural weather forecast RMSE by less than $1\%$, with negligible impact on 
the Fractions Skill Score (FSS).
\end{abstract}


\section{Introduction}


Climate simulations and reanalysis systems are continuously increasing in 
spatiotemporal resolution, and their storage demands are growing accordingly~\cite{nature2021,Bauer2015QuietRevolution,Overpeck2011ClimateDataChallenges}. 
This pressure is already evident in extreme-scale simulation workflows.
Recently, the ICON model~\cite{ICONModel,ICONSrc,ICONAMIPSetup} has been awarded the Gordon Bell Prize for Climate 
for high-resolution simulations at 1km spatial resolution and 10s temporal resolution~\cite{ICONGB}.
With a 3-hourly output frequency, it generates 16~TiB of data per checkpoint and
requires 200~GiB/s of storage write bandwidth and 620~GiB/s of read bandwidth, 
even in a configuration focused on showcasing the compute capability.
The Earth Virtualization Engines project~\cite{EVE} has reported up to 1~TB/s storage bandwidth requirements 
for high-resolution climate simulations.

These numbers show that storage and I/O are becoming 
primary bottlenecks in high-resolution climate simulations~\cite{EVE,ICONGB}.
Institutions around the world generate enormous volumes of climate data every day.
In one of the highest-resolution ICON runs~\cite{ICONGB}, 
the simulation sustained 145.7 simulated days per day,
meaning that a 3-hour simulation can be completed in 74.1s.
However, the I/O time needed to output a single time step at full spatial resolution
is 82.2s, which is longer than the simulation time itself.

This imbalance is problematic because it limits our ability to analyze 
the high-spatiotemporal-resolution data generated by these simulations.
To better understand the climate system, 
we need to be able to store and analyze the data produced by these simulations
at intervals much shorter than 3 hours.
That requires not only simulation models with higher computational performance
but also the ability to store and transmit much larger volumes of data.
The challenge is not limited to a single model.
In particular, the European Centre for Medium-Range Weather Forecasts (ECMWF) reported 
producing 230~TB of data per day in 2021, 
and next-generation systems are expected to approach the petabyte-per-day scale~\cite{nature2021,Bauer2015QuietRevolution}.
However, even for current-generation simulations, 
the required 1~TB/s of storage bandwidth is available at only a few supercomputers.
For example, Alps~\cite{AlpsStorage} provides 1.1~TB/s of write bandwidth.
Current workloads therefore place severe pressure on storage systems, I/O bandwidth, and data dissemination.

As supercomputers continue to become more powerful, this challenge becomes even greater,
because data can be generated faster than the available storage and I/O systems can handle.
We therefore need to reduce the amount of data that must be stored and transmitted
while preserving its quality for downstream applications.

\begin{figure}[t]
  \centering
  \includegraphics[width=\linewidth]{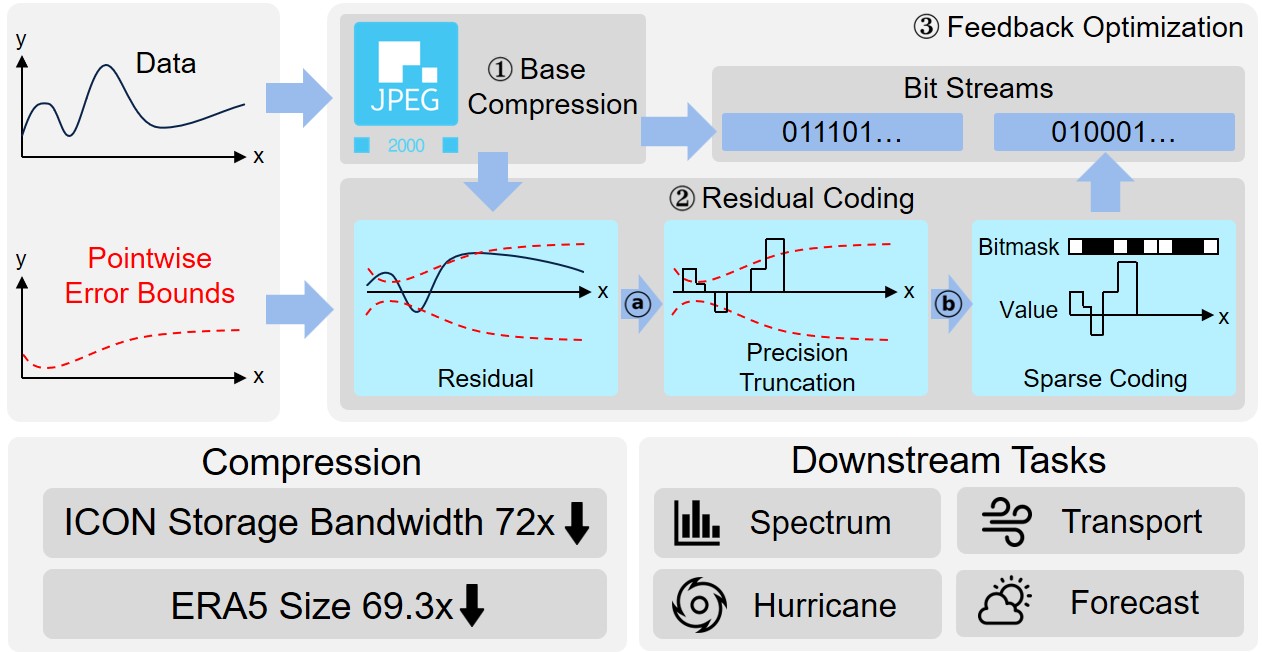}
  \caption{
    Overview of \projectname{}.
    \ding{172} - Base layer (JPEG2000 in our current implementation).
    \ding{173} - Residual coding step to ensure pointwise error bounds.
    \textcircled{a} - Error-bound-aware precision truncation by masking and rounding.
    \textcircled{b} - Lossless compression for the sparse representation of truncated residuals.
    \ding{174} - Optional feedback optimization to improve the overall compression ratio.
    In our evaluation, we show that \projectname{} achieves a high storage bandwidth reduction and compression ratio, while preserving data quality for downstream applications.
  }
  \label{fig:method_overview}
\end{figure}

Data compression is a promising solution to reduce storage and I/O costs.
Purely lossless compressors~\cite{lz4,zstd,xz} provide limited help at this scale.
For climate data, typical compression ratios are often around $2\times$~\cite{Kuhn2016},
which is valuable but insufficient when the primary bottleneck is dataset size.
Lossy compression can achieve compression ratios above $100\times$ on climate data, 
but its practical adoption remains limited. 
Without error control, lossy compression can introduce large distortions that affect downstream analyses. 
Although prior evaluations~\cite{CRA5,VideoCompressionEarthSystemData} suggest 
that some downstream tasks degrade only modestly at high compression ratios, 
users must be careful for unknown applications without explicit control of the compression error.

To address this issue, several error-bounded lossy compressors have been developed for scientific datasets~\cite{di2025survey,zfp,mgard, sz, sz3, sperr}.
Recent work has also proposed error-bounded designs tailored to climate data, including EBCC~\cite{ebcc}.
These methods provide strict guarantees on the maximum compression error.
However, in most existing workflows the tolerance remains a user-chosen global parameter,
typically expressed as an absolute, relative, or range-relative bound.
For a heterogeneous climate dataset such as ERA5~\cite{ERA5}, a single global tolerance
may be too strict in some regions and too permissive in others,
and appropriate values differ across variables and downstream tasks.
As a result, users must repeatedly tune and validate bounds for their own applications,
which makes it difficult to apply them in practice.

To make climate and weather data compression practical at scale, an ideal compression algorithm should satisfy four properties.
(1) The error guarantee should be directly linked to intrinsic properties of the data rather than to application-specific tuning.
(2) The algorithm should achieve a high compression ratio to efficiently reduce pressure on the I/O system and make the compressed dataset more portable.
(3) The algorithm should have high compression throughput to handle large data volumes with low resource usage.
(4) The compressed dataset should not introduce significant degradation in downstream applications.

\textit{Definition of statistically lossless compression.}
In this work, we propose statistically lossless compression that meets the desired properties for climate data compression.
In this new paradigm, instead of using artificially defined global error bounds, 
we define a compressed dataset to be \textit{statistically lossless} with respect to the original dataset
if the pointwise compression error is less than or equal to the pointwise uncertainty at every data point.
In other words, statistically lossless compression ensures that the compression error at each point does not exceed the intrinsic data uncertainty at that point.
Under this definition, compression does not move a value outside the pointwise uncertainty interval already represented by the dataset.

Statistically lossless compression requires two key components: 
a pointwise uncertainty estimate and a compressor that supports pointwise error bounds.
ERA5 includes a 10-member ensemble of data assimilations with inputs perturbed within the raw-measurement uncertainty range~\cite{ERA5,ERA5doc}.
The standard deviation of these ensemble members, i.e., the ensemble spread,
provides a pointwise uncertainty estimate for the final reanalysis fields~\cite{Leutbecher2008EnsembleForecasting,Dee2011ERAInterim}.
We use this ensemble spread as the pointwise error bound for compression.
To realize this idea, we propose \projectname{},
a two-stage compressor for floating-point climate data that enforces pointwise error bounds.
The first stage uses a rate-distortion-efficient base codec to exploit spatial structure,
and the second stage encodes the residuals to satisfy the pointwise error bounds while maximizing compression ratio.
An overview of our approach is shown in Fig.~\ref{fig:method_overview}.

In this paper, we demonstrate that our implementation of statistically lossless compression for climate data 
satisfies these four properties.
(1) In statistically lossless compression, the error target is directly tied to intrinsic properties of the data,
namely the uncertainty estimate provided by the climate dataset. 
The error bound is therefore not an arbitrary user choice but is grounded in the physical and statistical characteristics of the dataset.
(2) Our approach achieves a high compression ratio, 
reaching $37.7\times$ on the full ERA5 dataset and $69.3\times$ on the most representative variables.
The commonly used, most representative subset of ERA5 can be compressed to 4~TB and stored on a commodity hard drive.
Measured on real ICON output, our approach reduces I/O bandwidth requirements by $72\times$ and
can effectively increase the data output frequency from 3 hours to 7.5 minutes of simulated time at the same storage bandwidth.
(3) Our feedback-free design achieves a high compression throughput.
Measurements on real ICON output show that a single node provides
the compression throughput required for the data volumes reported
by current climate centers.
(4) We evaluate the compressed dataset on a comprehensive benchmark suite of downstream tasks,
where it achieves the highest compression ratio and a near-optimal rate-distortion trade-off among all evaluated methods,
while having only a limited impact on downstream applications.
We show that the compressed dataset preserves small-scale spectral features, extreme weather events,
and trajectory density, 
while it does not degrade weather forecast quality by more than $1\%$ in terms of RMSE with negligible impact on FSS.

\section{Method}

\subsection{Overview}


\projectname{} takes a floating-point array of shape $(N,H,W)$ and a pointwise error-bound array of the same shape.
The bounds are required only during compression and are not included in the compressed payload.
Its objective is to maximize compression ratio while satisfying every pointwise bound.
Fig.~\ref{fig:method_overview} shows the pipeline.

\ding{172} \textbf{Base layer.}
A spatial codec first compresses each $(H,W)$ slice independently.
The current implementation uses OpenJPEG~\cite{openjpeg,glymur} and JPEG2000~\cite{JPEG2000}, 
although any codec with a good rate-distortion trade-off can serve as the base layer.
A user-selected base-layer compression level controls the initial distortion.

\ding{173} \textbf{Residual coding.}
The base codec does not guarantee pointwise bounds, so \projectname{} encodes a correction for points that need one.
It has two substeps: precision truncation (\textcircled{a}) and sparse coding (\textcircled{b}).
Precision truncation retains only the residual bits required to satisfy each bound.
The resulting residuals contain many zeros and are stored using a bitmap and a packed array of nonzero values.
Both arrays are compressed losslessly with ZSTD~\cite{zstd}, and the value array is bit-shuffled before compression~\cite{cblosc,pythonblosc}.
Section~\ref{sec:residual_coding} gives the truncation rule and its guarantee.

\ding{174} \textbf{Feedback optimization.}
The residual layer guarantees the bounds for any base-layer compression level, 
but this parameter affects the combined size of the base and residual streams.
It can be selected in advance or optimized with golden-section search on a small sample.
Section~\ref{sec:feedback_optimization} provides more details on this component.

\textbf{Decompression.}
The decoder reconstructs the base layer and adds the decoded residuals to produce the reconstructed data.

\textbf{Floating-point mapping and final check.}
JPEG2000 operates on integers, so the compressor maps the data to uint16 by linearly scaling it to the uint16 range 
and scales the error bounds to the same domain.
The inverse mapping recovers floating-point values after decompression.
Rounding to uint16 introduces at most half a quantization step, approximately $8\times10^{-6}$ of the data range.
In typical scientific datasets, the pointwise bounds are $1\%$ of the data range, 
so this mapping error is at least three orders of magnitude smaller.
After inverse mapping, the compressor checks for any floating-point values that would violate their original bounds.
It stores the original values at those positions using the same lossless sparse representation 
and restores them during decompression.

\textbf{NaN handling.}
Some climate variables contain NaN values. 
For example, sea surface temperature is defined only over ocean grid points, 
and land points in both reanalysis and ensemble data are masked by NaN. 
\projectname{} preserves NaN values in the reconstruction by losslessly storing their locations in a bitmap.
Before compression, it replaces NaNs with 32768 and assigns their bounds the uint16 maximum, 
so these locations do not affect residual coding.
The decoder restores the NaNs from the bitmap.

\textbf{Bitstream composition and overhead.}
In our evaluation, sparse residual coding typically accounts for less than $5\%$ of the final bitstream.
For inputs containing NaNs, the NaN bitmap also accounts for less than $5\%$.
The final-check correction stream occupies 0 bytes because no violations occurred in our evaluation,
and metadata adds a constant 20 bytes to store the offsets of the bitstream sections.

\subsection{Error-Bound-Aware Precision Truncation}\label{sec:residual_coding}
Precision truncation (\textcircled{a}) of the residuals is the core component of \projectname{}.
Its primary role is to enforce the pointwise error bound.
At the same time, it is designed to maximize the compression ratio of the residual layer 
by zeroing out as many bits in the residuals as possible. 
In this section, we first show that the overall compression error is entirely determined by the precision truncation error, 
and then present the truncation rule and its guarantee.

\subsubsection{Why only precision truncation matters}
Let $u$ denote the original field in the integer domain and $\hat{u}$ the base-layer reconstruction.
The base-layer residual is defined as $r := u-\hat{u}$. Let $\hat{r}$ be its reconstruction after residual coding.
Then the final reconstruction is $\hat{u}' := \hat{u} + \hat{r}$.
By construction, 
the final compression error $Err$ equals the residual coding error:
\[
Err = u - \hat{u}' = u - (\hat{u}+\hat{r}) = r - \hat{r}
\]
Because the truncated residuals are stored losslessly in the sparse coding stage (\textcircled{b}),
this means that the final compression error is determined solely by the precision truncation step (\textcircled{a}).
Therefore, to guarantee the pointwise error bound, 
it is sufficient to guarantee the same bound for the precision truncation error.

\subsubsection{Precision truncation rule and guarantee}

\begin{figure}[th]
  \centering
  \includegraphics[width=\linewidth]{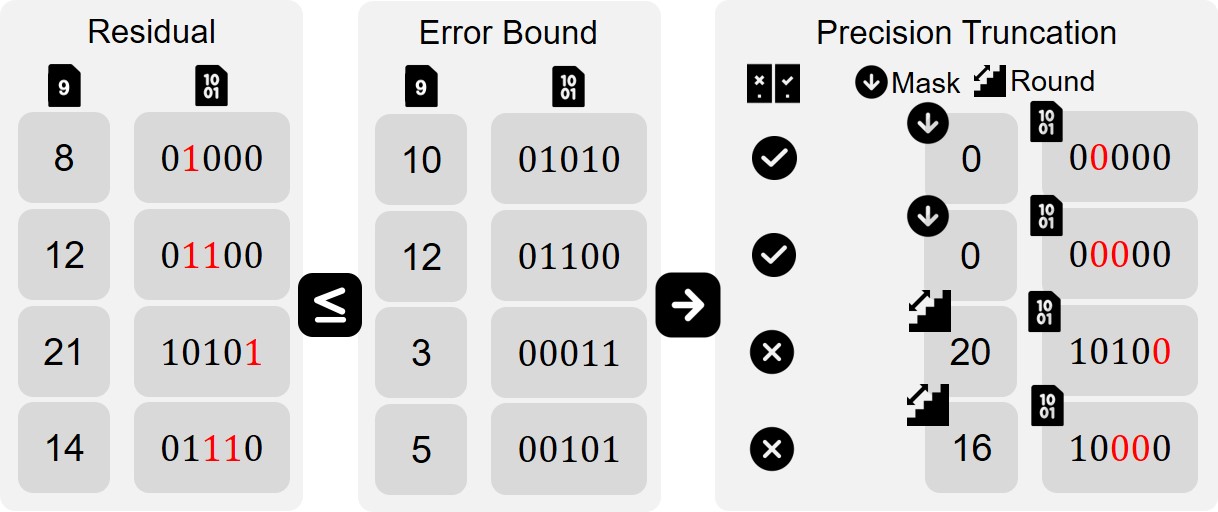}
  \caption{Precision truncation process (\textcircled{a}).
  The truncation error is never larger than the error bound.
  When the residual exceeds the error bound,
  apply rounding that maps the residual to a coarser lattice determined by the error bound.
  When the residual is within the error bound,
  apply masking to set the residual to zero.
  }
  \label{fig:residual_truncation}
\end{figure}

To guarantee the pointwise error bound while maximizing the compression ratio,
our precision truncation rule keeps only the most significant bits of the residual
that are accurate enough to satisfy the bound.
Fig.~\ref{fig:residual_truncation} illustrates this idea in the 8-bit case.

For each point, let $eb$ denote the error bound and $r$ denote the residual, and
define $S = 2^{\left\lfloor \log_2(2eb)\right\rfloor}$.
We then have the following precision truncation rule for the residual:
\[
\hat{r} =
\begin{cases}
\operatorname{RNE}(r/S)\,S, & |r|>eb,\\[4pt]
0, & |r|\le eb,
\end{cases}
\]
where $\operatorname{RNE}$ denotes round-to-nearest with ties-to-even.
This rule has two branches.

\textbf{Round.}
If the residual exceeds the pointwise bound,
we keep a quantized version of the residual.
The step size $S$ is chosen as the largest power of two no greater than $2eb$,
so rounding introduces no more than $eb$ error 
while zeroing as many least significant bits as possible.

\textbf{Mask.}
If the residual is already within the pointwise bound,
we set it to zero.
In this case, 
the base-layer reconstruction already satisfies the error bound at this point.
Therefore, we can set the residual to zero without violating the error bound.

Consequently, precision truncation ensures that 
the final reconstruction error remains within the pointwise error bound 
while trying to maximize the number of zero bits in the truncated residual
to achieve a high overall compression ratio.

\subsection{Feedback Optimization}\label{sec:feedback_optimization}

Although precision truncation guarantees that the final reconstruction satisfies the pointwise error bound 
for any base-layer compression level, the choice of this parameter can affect the overall compression ratio.
The overall compression ratio first increases and then decreases as the base layer becomes more aggressive.
Stronger base compression reduces the base stream but produces larger residuals. 
Weaker base compression has the opposite effect.
The resulting unimodal objective allows golden-section search to select a near-optimal level.
However, this search process requires multiple compression passes, which can be expensive for large datasets 
and lower the compression throughput.

To study the amount of data required for feedback optimization, 
we evaluate \projectname{} across four commonly used single-level variables from ERA5: 
10m u component (zonal) of wind, 10m v component (meridional) of wind, 2m temperature, and mean sea level pressure. 
The data are evaluated at full spatial resolution with 6-hour temporal intervals in December 2024, 
and the pointwise error bound is set equal to the ensemble spread.

\begin{figure}[th]
  \centering
  \includegraphics[width=\linewidth]{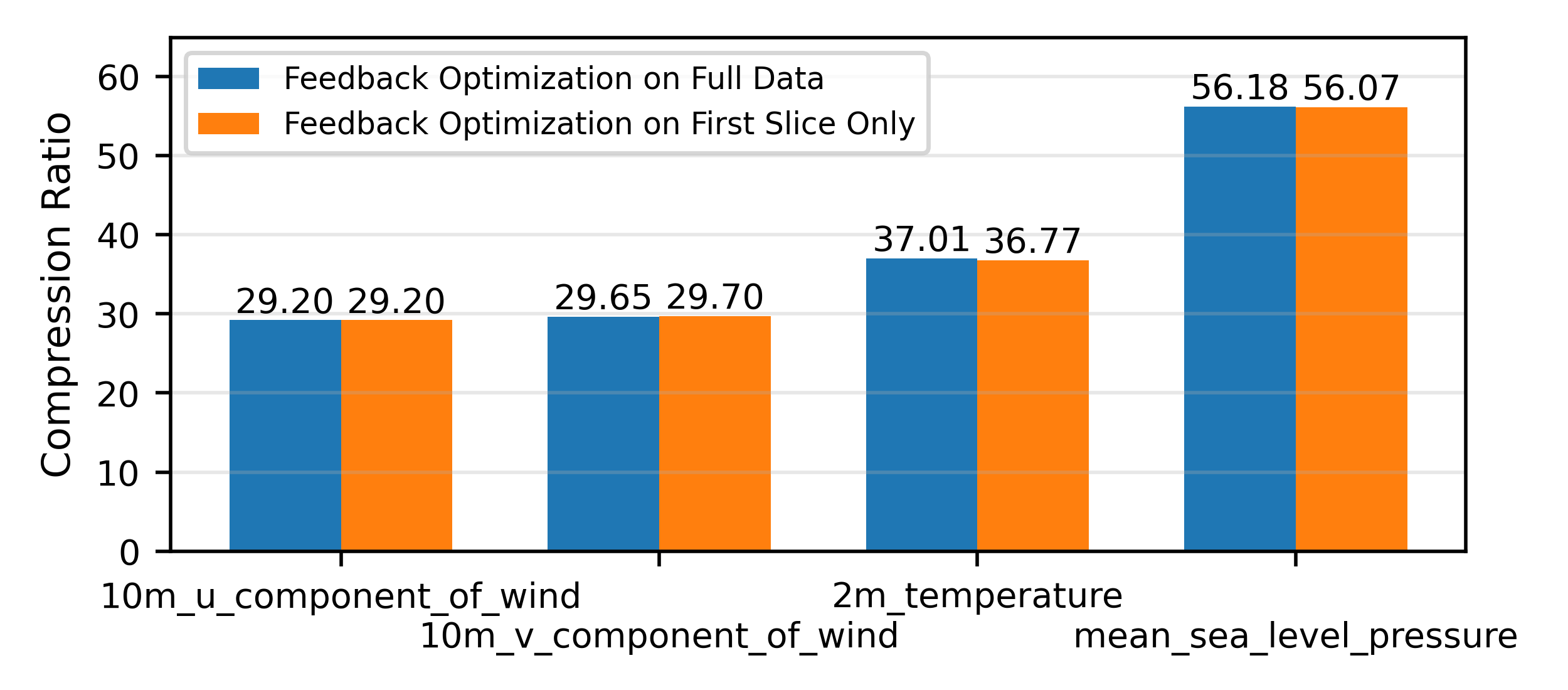}
  \caption{
    Compression-ratio difference when feedback optimization 
    uses the full input or only its first 2D slice.
  }
  \label{fig:ablation_study}
\end{figure}

Fig.~\ref{fig:ablation_study} shows that optimizing on the first slice 
changes the resulting compression ratio by less than $1\%$ across all four variables.
This suggests that feedback optimization can be run on a small subset of the data 
to reduce tuning cost without materially affecting compression throughput.

\section{Evaluation}

\subsection{Overview}

Compression ratio and throughput are central metrics for evaluating compression methods.
They directly reflect storage bandwidth reduction and computational efficiency.
However, assessing compression for climate data such as ERA5 requires more than 
reporting compression ratios and throughput for a given error target. 
Climate data is used for various downstream applications, including
visualization and analysis of extreme events, 
tracing atmospheric transport, and evaluating forecasting models.
These applications are sensitive to different forms of reconstruction error,
and a good compression method should have limited impact on all of these scientific uses of the data.

To give a comprehensive evaluation of \projectname{}, 
we therefore combine algorithm- and application-level validation in our evaluation,
following prior work~\cite{ebcc,Baker2014,ZChecker2017}.
The following evaluations are organized around six questions.
(1) Can \projectname{} efficiently enforce statistically lossless compression under pointwise uncertainty bounds?
(2) Can its compression performance translate into substantial storage-bandwidth reduction for real weather-simulation output?
(3) Under conventional global error bounds, how does \projectname{} compare to existing error-bounded compressors?
(4) Does the compressed dataset produced by \projectname{} introduce artifacts that affect downstream applications?
(5) Does our approach remain effective when scaled from the most representative variables to all variables in ERA5?
(6) How robust is \projectname{} to changes in the supplied uncertainty estimates?

\subsection{Statistically Lossless Compression of the Most Representative ERA5 Variables}\label{sec:pointwise_common}

\begin{figure}[th]
  \centering
  \includegraphics[width=\linewidth]{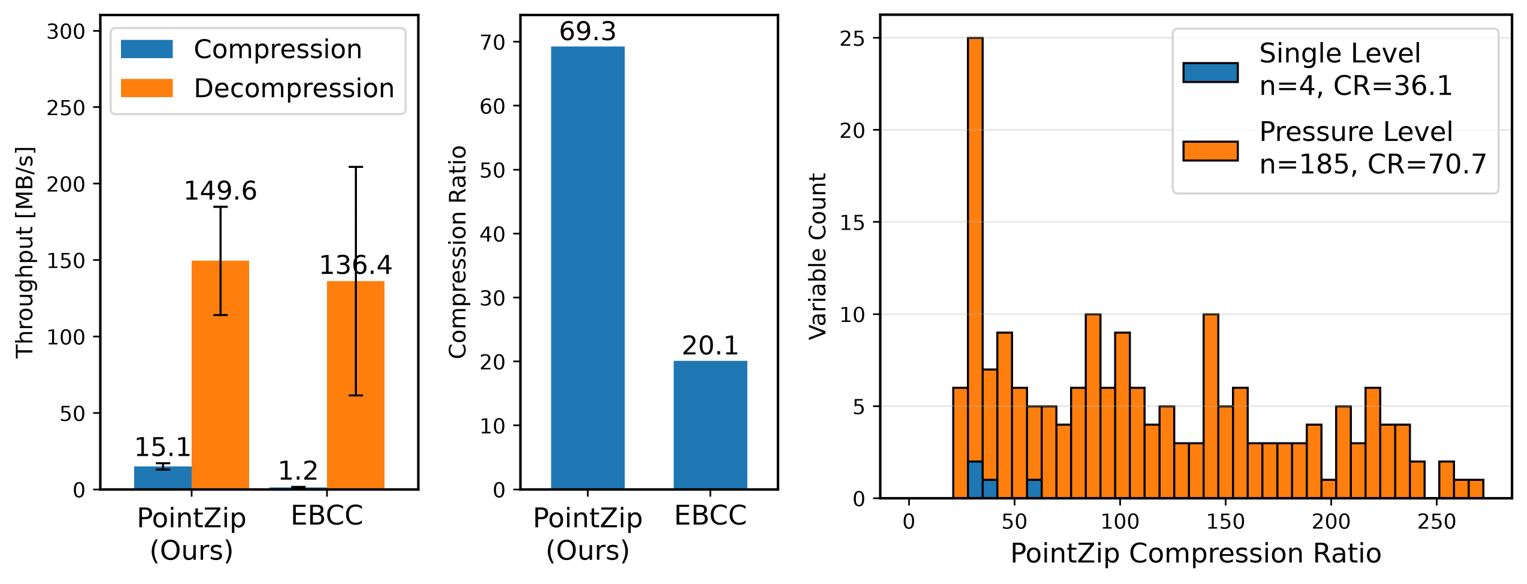}
  \caption{
    Statistically lossless compression results for the most representative ERA5 variables.
    Left: single-threaded compression and decompression throughput.
    Error bars denote standard deviation.
    Middle: overall compression ratio for different compressors.
    Right: histogram of \projectname{} compression ratios across variables and pressure levels.
  }
  \label{fig:pointwise_common}
\end{figure}

This section answers question (1) by evaluating \projectname{} under the main setting of this work,
namely statistically lossless compression under pointwise error bounds defined by the ERA5 ensemble spread.

\subsubsection{Evaluation setup}



We evaluate December 2024 ERA5 data at hourly resolution on the global $0.25^\circ$ grid.
Each variable and pressure level is compressed as a separate chunk containing 744 time steps and $721\times1440$ spatial points.
The lower-resolution ensemble spread is linearly interpolated to the reanalysis grid and used as the pointwise error bound.

To simulate realistic usage scenarios, 
we focus on the most representative subset of ERA5 variables.
This subset contains four single-level variables, 
2m temperature, 10m u component of wind, 10m v component of wind, and mean sea level pressure, 
and five pressure-level variables, temperature, u component of wind, v component of wind, specific humidity, and geopotential, 
on all 37 pressure levels.
They are the main variables used in weather and climate applications, 
including both traditional physics-based analysis~\cite{lagranto} and 
neural weather modeling~\cite{pangu,graphcast,aurora}.
The raw data size of this subset is 584~GB in our evaluation period,
and our performance measurements are performed on an AMD EPYC 7742 CPU.

No baseline compressor natively accepts pointwise error bounds as constraints.
EBCC~\cite{ebcc} is the most relevant compression method as it is designed specifically for weather and climate data.
It uses a feedback loop to adjust its compression parameters until all points satisfy the error target.
We therefore extend EBCC to support pointwise error bounds
by modifying its feedback loop so that tuning stops only when all points satisfy their pointwise bounds.
This extended version of EBCC is the baseline method for comparison in this section.

\subsubsection{Results}

Fig.~\ref{fig:pointwise_common} summarizes the results.
\projectname{} has 15.1~MB/s single-threaded compression throughput and
149.6~MB/s single-threaded decompression throughput.
Its compression throughput is $12.6\times$ higher than that of EBCC,
while decompression throughput is of the same order of magnitude.
This aligns with the design of the method.
\projectname{} uses predefined compression parameters and a single-pass encoding process,
whereas EBCC must repeatedly recompress inside a feedback loop until pointwise constraints are met.

Besides high compression throughput,
\projectname{} has an overall compression ratio of $69.3\times$ on the full subset. 
This is $3.4\times$ higher than that of EBCC under the same pointwise error bounds.
When admissible error varies strongly across space,
a method that does not explicitly isolate the hard points is effectively limited by the tightest tolerances.
In contrast, \projectname{} handles those points by a residual coding step, 
which allows it to achieve a much higher compression ratio while still satisfying the pointwise error bounds.

Across individual variables, 
\projectname{} achieves compression ratios between $25\times$ and $60\times$ for the single-level variables, 
and between $25\times$ and $270\times$ for the pressure-level variables. 
This broad range reflects the strong variation in 
compressibility across variables and pressure levels,
but \projectname{} consistently delivers substantial size reduction throughout the benchmark.

To put those results in an operational context, 
applications are mainly interested in the recent 40 years of
the most representative ERA5 variables, which have a total size of 275~TB.
Our method can reduce this to 4~TB in a statistically lossless manner,
which can be stored on a single commodity hard drive or transferred over the network quickly.
This makes it possible to distribute a single compressed version of the dataset that is broadly usable.

\subsection{Storage Bandwidth Reduction for Weather Simulation}\label{sec:icon_compression}

\begin{figure}[th]
  \centering
  \includegraphics[width=\linewidth]{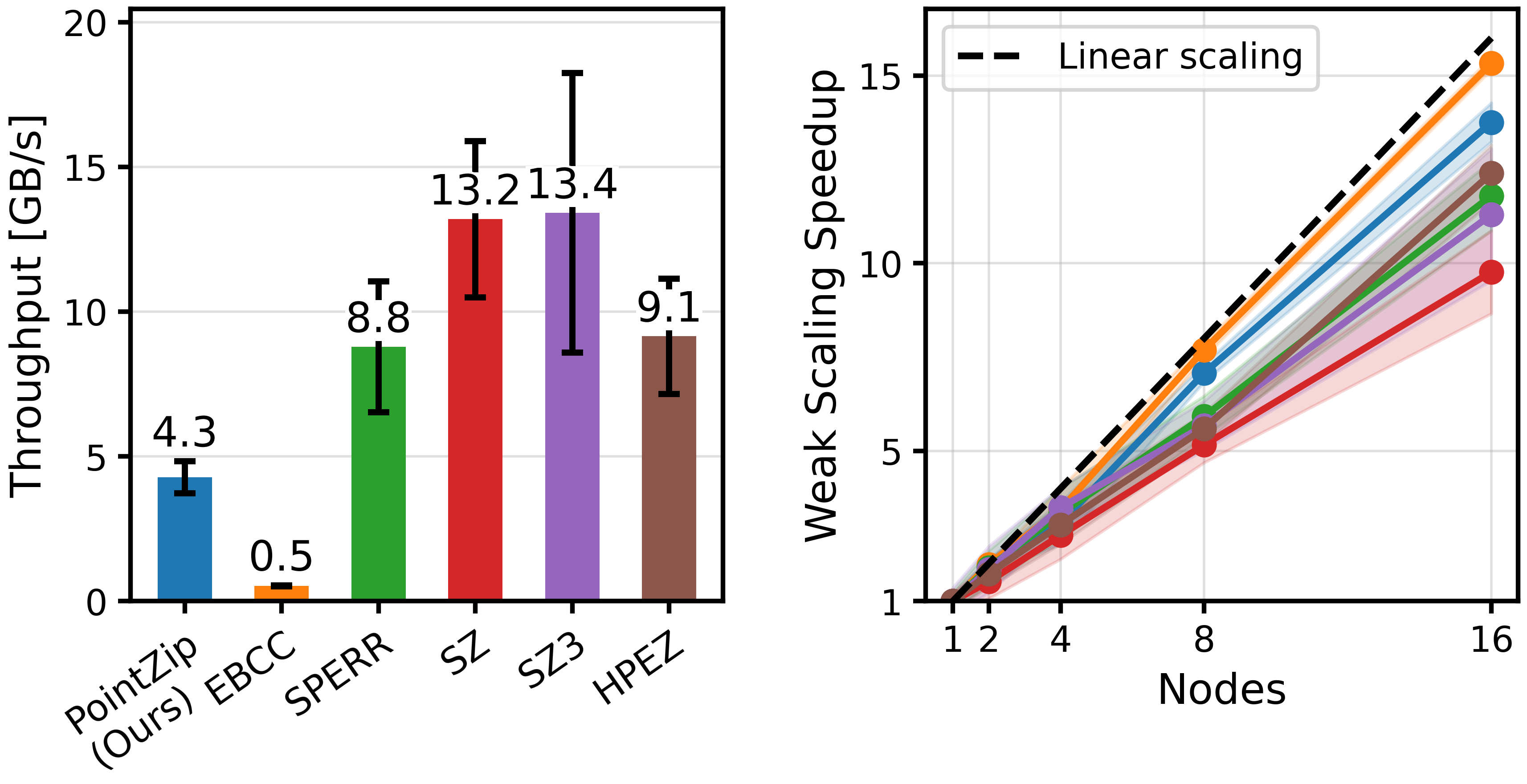}
  \caption{
    Multi-threaded compression performance on real ICON output. 
    Left: single-node compression throughput. 
    Right: weak-scaling speedup from 1 to 16 nodes. The dashed line denotes ideal linear scaling.
  }
  \label{fig:icon_compression}
\end{figure}

This section answers question (2) by evaluating \projectname{} on real output from the ICON weather model~\cite{ICONModel,ICONSrc,nextGEMS},
and analyzing the implications of our method for storage bandwidth in weather-simulation workflows.

\subsubsection{Evaluation setup}

To analyze the implications for storage bandwidth in weather simulation,
we apply our method to raw output from the ICON model~\cite{ICONModel,ICONSrc,nextGEMS}.
ICON is widely used in major weather centers and 
is capable of global modeling at 1km resolution~\cite{ICONGB}.
We use the output of a 16-step air bubble simulation example provided by ICON in the default setup, 
with results remapped to a $0.25^\circ$ rectangular grid.
Pointwise error bounds are set based on the model discretization error,
which is estimated by measuring the difference in results when doubling the spatial and temporal resolution.

In addition to the baseline method in Section~\ref{sec:pointwise_common},
we also compare to other state-of-the-art compressors that only provide global error bounds,
including SPERR~\cite{sperr}, SZ~\cite{sz}, SZ3~\cite{sz3}, and HPEZ~\cite{HPEZ}.
We also exclude the constant fields in the data because they can be compressed with a simple constant-value encoding and
do not reflect the general performance of the method on floating-point data.
Each node uses 288 threads to compress all variables and pressure levels. Different nodes compress different time steps in parallel.
The raw data size is 42.7~GB per node in our evaluation period, matching the load per GH200 node of the 1.25km resolution climate simulation~\cite{ICONGB}.

\subsubsection{Results}

Under the pointwise error bounds derived from the ICON discretization error, \projectname{} achieves a compression ratio of $108\times$, compared with $13.2\times$ for the pointwise-bound extension of EBCC.
The left panel of Fig.~\ref{fig:icon_compression} shows multi-threaded compression throughput on a single node with 288 threads.
\projectname{} reaches 4.3~GB/s, corresponding to approximately 372~TB/day.
This exceeds the 230~TB/day data-production rate reported by ECMWF~\cite{nature2021}, indicating that a single node can sustain an operational-scale compression workload.
EBCC reaches only 0.5~GB/s under the same pointwise constraints and is therefore $8.6\times$ slower.
SPERR, SZ, SZ3, and HPEZ achieve higher raw compression throughput, ranging from 8.8 to 13.4~GB/s, but do not provide the spatially varying pointwise guarantees evaluated here.
The right panel shows that \projectname{} achieves approximately $14\times$ weak-scaling speedup on 16 nodes.
This demonstrates that its throughput scales effectively as the data volume and number of compression nodes increase.

\begin{figure}[t]
  \centering
  \includegraphics[width=\linewidth]{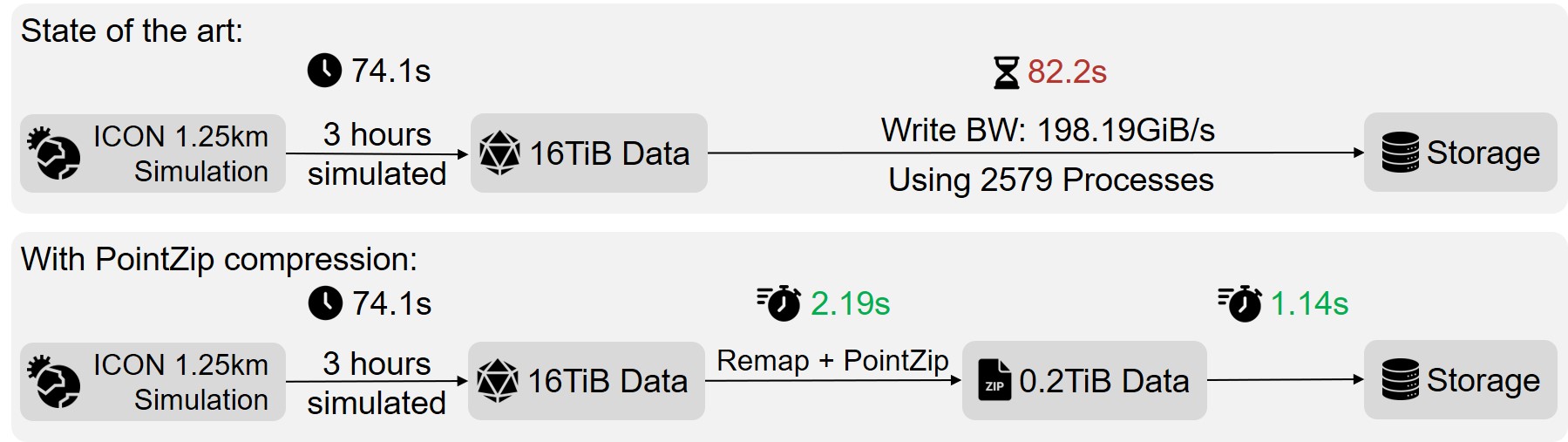}
  \caption{
    Projected impact of \projectname{} on the 1.25~km ICON output workflow.
  }
  \label{fig:icon_projection}
\end{figure}

We project our results to a large-scale weather-simulation workflow
by adopting numerical values from the Gordon Bell Prize run~\cite{ICONGB}.
As shown in Fig.~\ref{fig:icon_projection},
this run uses 2579 processes to output 16~TiB of data at the end of the 3-hour simulation.
To apply our method, we first remap the data to a rectangular grid, then apply \projectname{}
using 288 threads per process.
This grid mapping step increases the data size by $50\%$,
so our method achieves $72\times$ overall compression on the original data, thus reducing the time to output the data to storage from 82.2s to 3.33s.

This $24.7\times$ reduction in end-to-end output time can alternatively allow for an increase in the output frequency
from 3 hours to 7.5 minutes of simulated time without increasing the storage bandwidth or runtime requirements.
This timescale is important for post hoc air parcel tracing and for precipitation and convection lifecycle analysis.
Under the same output frequency, the bandwidth requirement is reduced from 198.19~GiB/s to 2.75~GiB/s,
lowering the required storage bandwidth by $72\times$ 
and enabling the use of commodity storage systems for global 1.25km resolution climate simulations.

\subsection{Compression under Global Error Bounds of the Most Representative ERA5 Variables}\label{sec:maxerror_common}


\begin{figure}[th]
  \centering
  \includegraphics[width=0.9\linewidth]{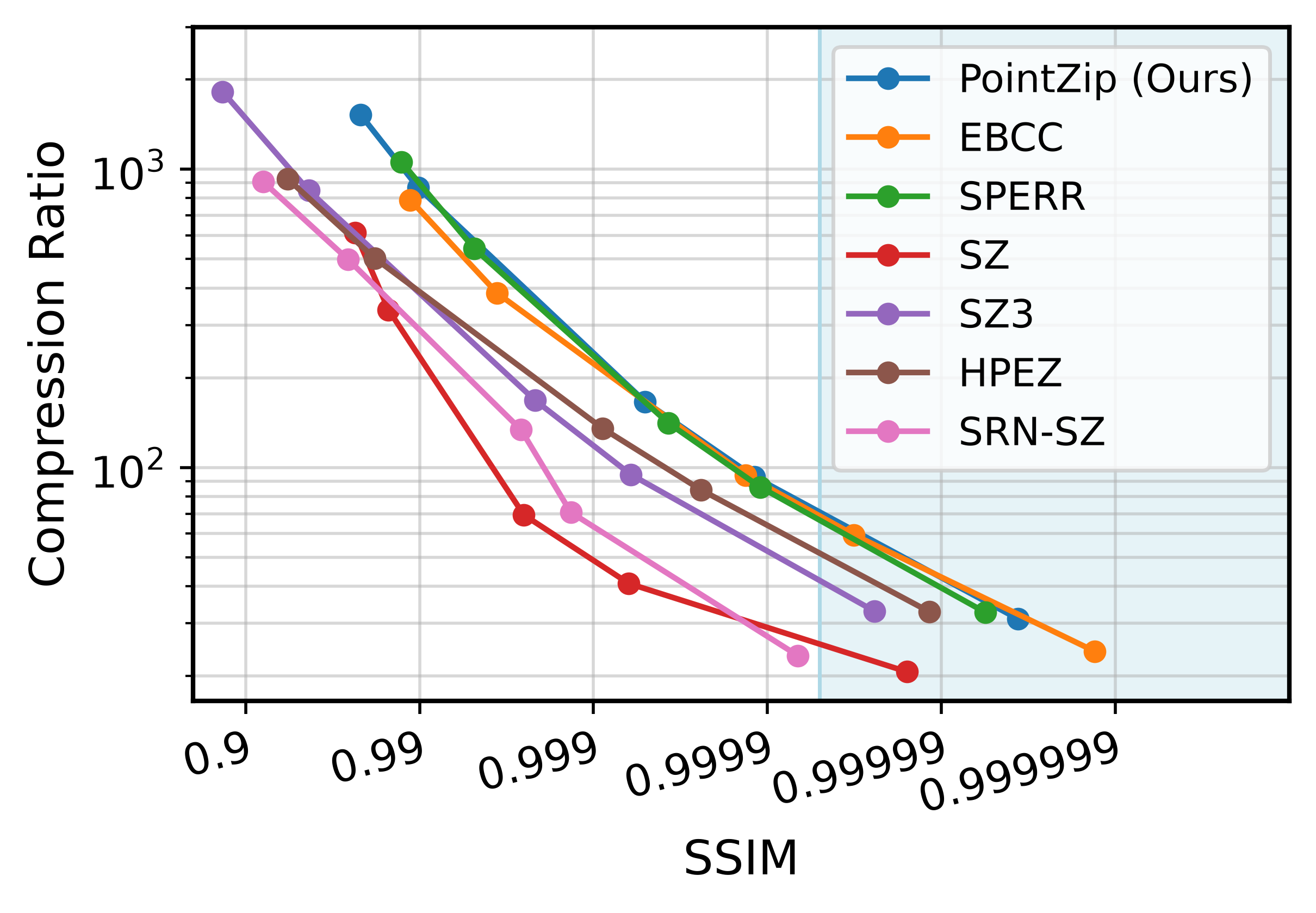}
  \caption{
    Rate-distortion trade-off, compression ratio vs. SSIM. 
    The 0.99995 SSIM threshold proposed by Baker et al.~\cite{Baker2014} is marked by the blue region on the plot.
  }
  \label{fig:maxerror_common}
\end{figure}

This section answers question (3) by evaluating \projectname{} under global error bounds
and comparing it to existing error-bounded compressors.

\subsubsection{Evaluation setup}
We use the same data and measurement setup as in Section~\ref{sec:pointwise_common},
except that all methods are evaluated under global error bounds.
However, the choice of global error bound is usually application-dependent in practice, 
and it is not meaningful to set a single global error bound across variables with different units and dynamic ranges.
We therefore use a range-relative error bound that is defined
as a constant fraction of the data range (maximum value minus minimum value) of each data chunk.
We evaluate fractions of $0.1\%, 0.5\%, 1\%, 5\%, 10\%$ to span strict, moderate, and relaxed error regimes.
This provides a consistent basis across variables and levels.

In addition to the baseline method in Section~\ref{sec:icon_compression},
we also include SRN-SZ~\cite{srn-sz} as our baseline.
SRN-SZ is a recent neural-network-based compressor built on top of SZ3.
We train SRN-SZ on ERA5 data using the official training scripts.

\subsubsection{Results}


We evaluate the rate-distortion trade-off of different compressors by measuring 
the compression ratio and the structural similarity index (SSIM) of the reconstructed fields
at different error bounds. 
Fig.~\ref{fig:maxerror_common} summarizes the results.
Following Baker et al.~\cite{Baker2014}, 
we consider compressed fields with SSIM values above 0.99995 to be visually indistinguishable from the original fields.
This region is highlighted in blue in the plot.
\projectname{} provides the strongest rate-distortion trade-offs 
together with SPERR and EBCC.
This means that even though \projectname{} is not designed to optimize for global error bounds,
it still achieves high compression ratios while maintaining high visual fidelity under conventional settings.



\subsection{Downstream Applications at Matched Compression Ratios}\label{sec:matched_compression_downstream}

This section answers question (4) by evaluating the compressed datasets 
in a wide range of downstream applications including both 
traditional physics-based applications and modern AI-based applications.

To compare the impact of compression from different compressors,
we evaluate the compressed datasets under the same storage budget.
We chunk data in the same way as in Section~\ref{sec:pointwise_common} and
compress it under a statistically lossless setup with \projectname{}.
For the same set of baseline methods as in Section~\ref{sec:maxerror_common}, 
we tune their constant global error bounds with a binary search
until their overall compression ratio matches the \projectname{} target within $\pm 1$.

These compressed datasets at matched compression ratios are then 
evaluated in a comprehensive benchmark suite of downstream analyses.
Those analyses include general analysis of the data spectrum, visual analysis of extreme events, 
traditional physics-based Lagrangian trajectory simulations, and weather forecasting by neural networks.

\subsubsection{Spectrum analysis}

\begin{figure}[th]
  \centering
  \includegraphics[width=\linewidth]{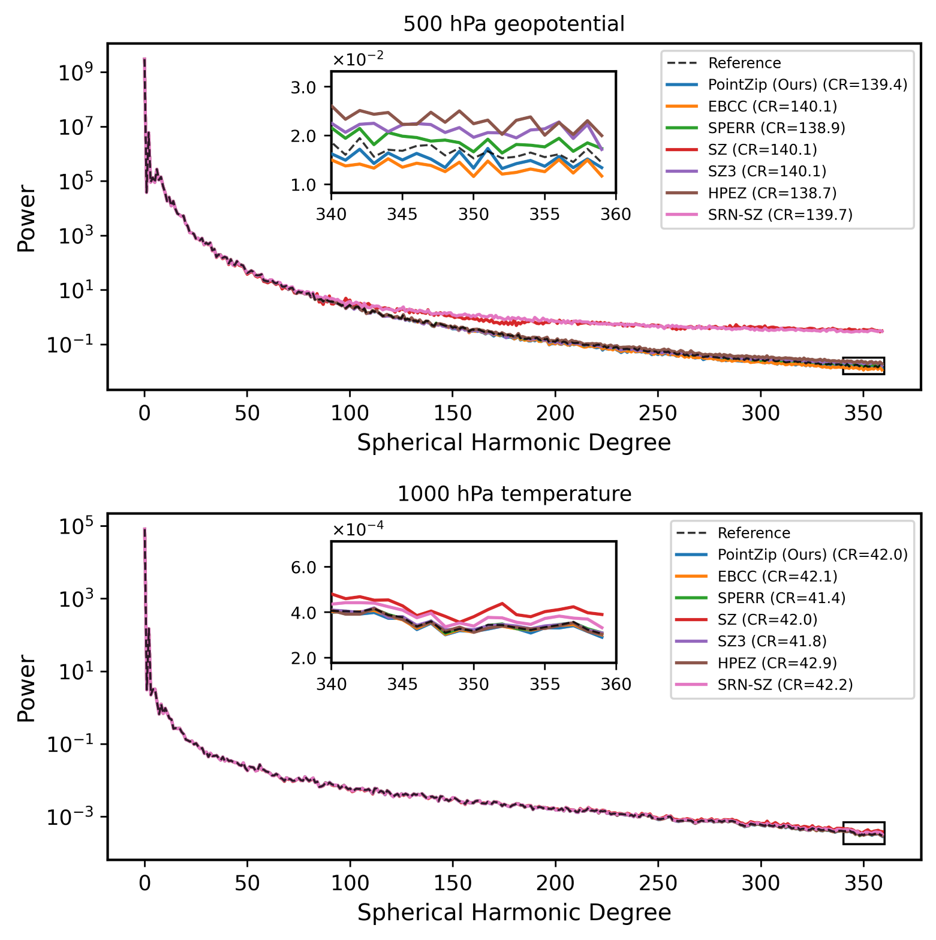}
  \caption{
    Spherical-harmonics power spectra under matched compression ratios.
    Top: geopotential (\texttt{z}) at 500~hPa.
    Bottom: temperature (\texttt{t}) at 1000~hPa.
    Insets show zoomed views of the high-frequency range.
  }
  \label{fig:spectrum_analysis}
\end{figure}

\begin{figure*}[th]
  \centering
  \includegraphics[width=\linewidth]{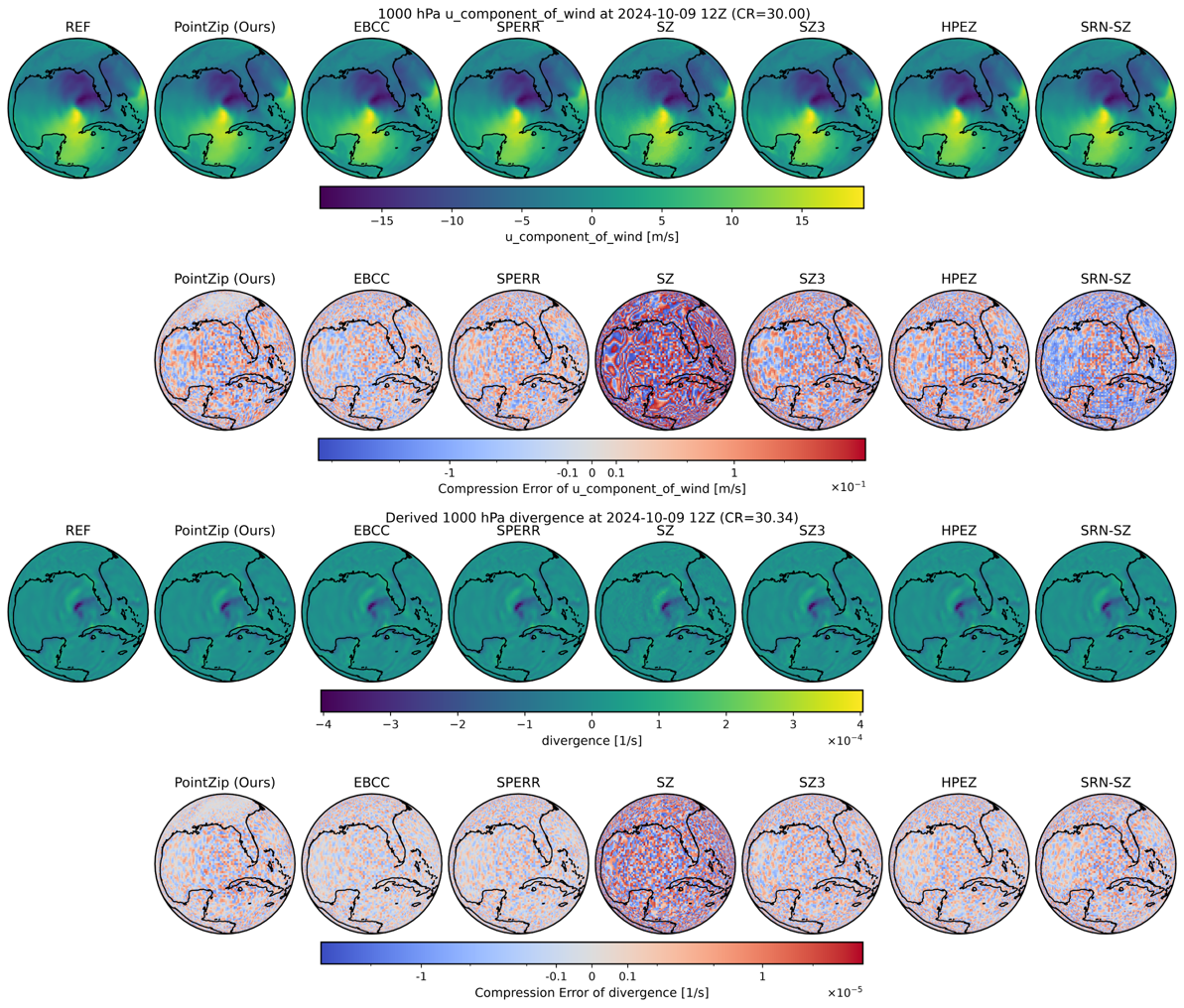}
  \caption{
    Synthesized satellite views of wind and wind-derived fields during Hurricane Milton at 1000~hPa under matched compression ratios.
    Top: u component of wind.
    Bottom: divergence derived from the compressed wind fields.
    For each derived field, the upper row shows the reconstructed field and the lower row shows the error relative to the reference derived from the original wind fields.
  }
  \label{fig:milton_1000hPa}
\end{figure*}

\begin{figure*}[th]
  \centering
  \includegraphics[width=\linewidth]{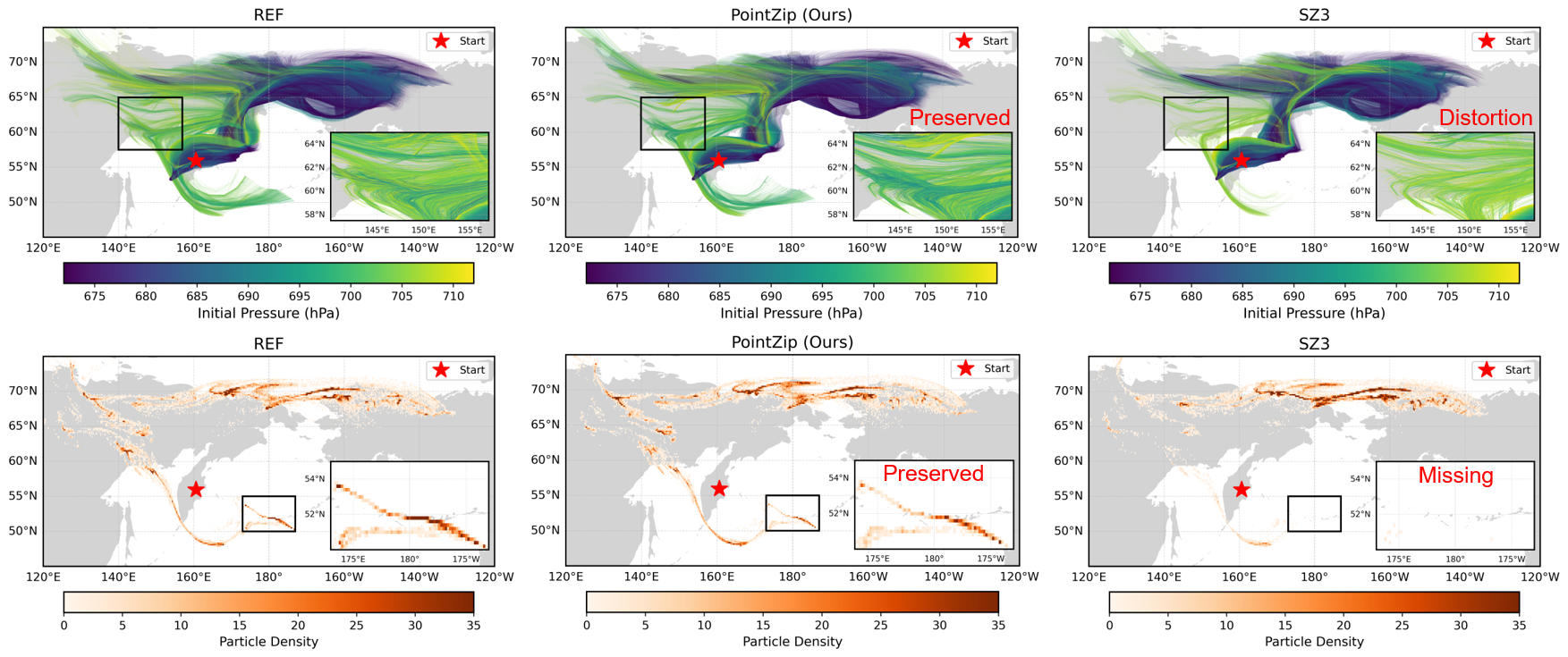}
  \caption{
    Seven-day air-parcel trajectories under matched compression ratios for the Bezymianny eruption.
    Trajectories are initialized from 50,000 locations near the top of the volcano.
    Top: all trajectories, colored by initial pressure level.
    Bottom: particle-density maps at the end of the seven-day period with a resolution of $0.2^\circ$.
  }
  \label{fig:trajectory}
\end{figure*}

To examine scale-dependent fidelity of different compressors at matched compression ratios, 
we compute the spherical-harmonics power spectra of the compressed fields and compare them to the original spectra.
Fig.~\ref{fig:spectrum_analysis} shows two representative examples. 
Even at a moderate compression ratio ($42.0 \times$), SZ and SRN-SZ deviate visibly from the reference spectrum 
in the high-frequency range for temperature at 1000~hPa. 
At a higher compression ratio ($139.4 \times$), SPERR, EBCC and \projectname{} remain close to the original spectrum,
whereas other methods introduce larger artifacts in the high-frequency range.

These spectral differences matter because high-frequency components encode sharp gradients 
and small-scale structures that can affect downstream analyses.
The spectral results therefore suggest that \projectname{} preserves physically relevant fine-scale structures rather than smoothing them away.



\subsubsection{Case study of Hurricane Milton}

High-frequency artifacts are visible not only in spectra but also directly in spatial fields around extreme events. 
Here we present a case study of Hurricane Milton, which formed in the Gulf of Mexico in October 2024 and reached Category 5 intensity.
Fig.~\ref{fig:milton_1000hPa} shows 
synthetic satellite views of the storm and the derived divergence field at 1000~hPa.
\projectname{}, EBCC and SPERR preserve the storm structure 
with better quality than other methods. On the divergence field,
\projectname{} shows slightly higher reconstruction error than EBCC and SPERR,
but still maintains a higher quality than SZ, SZ3, HPEZ and SRN-SZ.
This result is consistent with the spectrum analysis and shows 
that \projectname{} preserves 
the structure of extreme-weather events rather than smoothing it away.

\subsubsection{Case study of Bezymianny volcano eruption}

\begin{table}[th]
\centering
\caption{Air parcel density FSS $\uparrow$ vs. time after eruption.}
\label{tab:density_fss}
\begin{tabular}{lcccc}
\toprule
Method & 1 day & 3 days & 5 days & 7 days \\
\midrule
PointZip (Ours) & \underline{0.994} & \underline{0.907} & 0.843 & \underline{0.804} \\
EBCC            & \textbf{0.997} & \textbf{0.963} & \textbf{0.945} & \textbf{0.875} \\
SPERR           & 0.873 & 0.890 & 0.876 & 0.791 \\
SZ              & 0.288 & 0.000 & 0.000 & 0.017 \\
SZ3             & 0.626 & 0.296 & 0.311 & 0.335 \\
HPEZ            & 0.984 & 0.895 & \underline{0.895} & 0.797 \\
SRN-SZ          & 0.927 & 0.619 & 0.219 & 0.258 \\
\bottomrule
\end{tabular}
\end{table}

Traditional physics-based applications are also important downstream uses of ERA5.
Here we evaluate the effect of compression on Lagrangian trajectory simulation.
We present a case study of the Bezymianny volcano eruption on December 24, 2024.
Volcanic eruptions inject material into the atmosphere,
and the subsequent transport of this material is important
for understanding the eruption's impact on air quality and for aviation safety.
This is a challenging application for compression because errors in the wind fields can
accumulate over time and lead to large deviations in the simulated trajectories,
so it is important to understand how compression affects the results.

We run a seven-day simulation with LAGRANTO~\cite{lagranto} using compressed datasets under matched compression ratios. 
A total of 50,000 air parcels are initialized near the volcano summit.
To compare, we compute the particle density field at the end of each day 
and evaluate the Fractions Skill Score (FSS) of the final density field relative 
to the reference density field obtained from the original wind fields.
The FSS measures the spatial agreement between neighborhood event fractions in reconstructed and reference fields. It ranges from 0 to 1, with higher values indicating better agreement.
EBCC is closest to the reference.
\projectname{}, HPEZ and SPERR are slightly worse but still maintain a higher FSS than other methods
(Table~\ref{tab:density_fss}).

Fig.~\ref{fig:trajectory} shows the simulated trajectories and 
the final particle-density maps at the end of the seven-day period with a resolution of $0.2^\circ$.
For high-fidelity methods including \projectname{},
the details in the trajectories are preserved and lead to a more accurate density field.

\subsubsection{Weather Forecast Impact}

\begin{table*}[th]
\centering
\caption{Aurora forecast performance across 6-hour rollout steps under matched compression ratios.}
\label{tab:forecast_rollout_steps}
\setlength{\tabcolsep}{4pt}
\renewcommand{\arraystretch}{1.08}
\scriptsize
\resizebox{\textwidth}{!}{%
\begin{tabular}{lcccccccc}
\toprule
& \multicolumn{2}{c}{Step 1}
& \multicolumn{2}{c}{Step 2}
& \multicolumn{2}{c}{Step 3}
& \multicolumn{2}{c}{Step 4} \\
\cmidrule(lr){2-3}
\cmidrule(lr){4-5}
\cmidrule(lr){6-7}
\cmidrule(lr){8-9}
Method
& RMSE $\downarrow$ & FSS $\uparrow$
& RMSE $\downarrow$ & FSS $\uparrow$
& RMSE $\downarrow$ & FSS $\uparrow$
& RMSE $\downarrow$ & FSS $\uparrow$ \\

\midrule

REF
& 0.00745
& 0.9980
& 0.01044
& 0.9962
& 0.01264
& 0.9943
& 0.01458
& 0.9924 \\

\midrule

PointZip (Ours)
& \underline{0.00748 (+0.40\%)}
& \textbf{0.9980 (-0.00\%)}
& \textbf{0.01045 (+0.10\%)}
& \textbf{0.9962 (-0.00\%)}
& \textbf{0.01265 (+0.08\%)}
& \textbf{0.9943 (-0.00\%)}
& \textbf{0.01459 (+0.07\%)}
& \textbf{0.9924 (-0.00\%)} \\

EBCC
& \textbf{0.00747 (+0.27\%)}
& \textbf{0.9980 (-0.00\%)}
& \textbf{0.01045 (+0.10\%)}
& \textbf{0.9962 (-0.00\%)}
& \textbf{0.01265 (+0.08\%)}
& \textbf{0.9943 (-0.00\%)}
& \textbf{0.01459 (+0.07\%)}
& \textbf{0.9924 (-0.00\%)} \\

SPERR
& 0.00775 (+4.03\%)
& \underline{0.9979 (-0.01\%)}
& \underline{0.01057 (+1.25\%)}
& \underline{0.9961 (-0.01\%)}
& \underline{0.01272 (+0.63\%)}
& \textbf{0.9943 (-0.00\%)}
& \underline{0.01463 (+0.34\%)}
& \textbf{0.9924 (-0.00\%)} \\

SZ
& 0.01617 (+117\%)
& 0.9898 (-0.82\%)
& 0.01751 (+67.7\%)
& 0.9886 (-0.76\%)
& 0.01892 (+49.7\%)
& 0.9866 (-0.77\%)
& 0.02030 (+39.2\%)
& 0.9844 (-0.81\%) \\

SZ3
& 0.01396 (+87.4\%)
& 0.9927 (-0.53\%)
& 0.01524 (+46.0\%)
& 0.9919 (-0.43\%)
& 0.01639 (+29.7\%)
& 0.9906 (-0.37\%)
& 0.01764 (+21.0\%)
& 0.9891 (-0.33\%) \\

HPEZ
& \underline{0.00748 (+0.40\%)}
& \textbf{0.9980 (-0.00\%)}
& \textbf{0.01045 (+0.10\%)}
& \textbf{0.9962 (-0.00\%)}
& \textbf{0.01265 (+0.08\%)}
& \textbf{0.9943 (-0.00\%)}
& \textbf{0.01459 (+0.07\%)}
& \textbf{0.9924 (-0.00\%)} \\

SRN-SZ
& 0.00819 (+9.93\%)
& 0.9978 (-0.02\%)
& 0.01093 (+4.69\%)
& 0.9960 (-0.02\%)
& 0.01303 (+3.09\%)
& \underline{0.9942 (-0.01\%)}
& 0.01491 (+2.26\%)
& \underline{0.9923 (-0.01\%)} \\

\bottomrule
\end{tabular}%
}
\vspace{-0.25cm}
\end{table*}

Modern AI-based applications are increasingly important downstream uses of ERA5.
One typical example is neural weather modeling,
so it is important to understand how compression affects forecast skill. 
We evaluate this using Aurora~\cite{aurora}, a neural foundation model for Earth-system prediction,
and measure forecast skill when compressed ERA5 data are used as input.
Predictions are based on compressed ERA5 data and 
are fed back autoregressively for four steps at 6-hour intervals, corresponding to 24 hours.

Table~\ref{tab:forecast_rollout_steps} shows the 
normalized latitude-weighted root mean squared error (RMSE) and 
the Fractions Skill Score (FSS) of Aurora rollouts.
\projectname{} changes forecast skill only marginally.
Across all lead times, the RMSE increase remains below 1\% and the FSS decrease is negligible.

\subsection{Statistically Lossless Compression of all ERA5 Variables}\label{sec:pointwise_all}

This section answers question (5) by evaluating statistically lossless compression of
all ERA5 variables.

\begin{table}[thbp]
\centering
\caption{Compression summary for all ERA5 variables.}
\label{tab:era5_summary}
\begin{tabular}{lrrr}
\toprule
\textbf{Level} & \textbf{Single} & \textbf{Pressure} & \textbf{Total} \\
\midrule
Variables & 262 & 592 & 854 \\
\midrule
\multicolumn{4}{l}{\textbf{Variables with Special Values}} \\
NaN-Padding & 50 & 0 & 50 \\
Zero in Ensemble Spread & 143 & 185 & 328 \\
\midrule
\multicolumn{4}{l}{\textbf{Skipped Variables}} \\
Missing Ensemble Spread File & 36 & 0 & 36 \\
Constant Boundary Conditions & 22 & 0 & 22 \\
\midrule
\multicolumn{4}{l}{\textbf{Compression Results}} \\
Variables & 204 & 592 & 796 \\
Uncompressed Size [TB] & 0.52 & 1.83 & 2.35 \\
Compression Ratio & 23.2 & 45.9 & 37.7 \\
\bottomrule
\end{tabular}
\end{table}

\subsubsection{Evaluation setup}
We apply statistically lossless compression to all ERA5 variables
under the same chunking setup and time period as in Section~\ref{sec:pointwise_common}.
ERA5 contains both single-level and pressure-level variables,
with a total of 854 variables across different levels.
50 of these variables contain NaN values for padding,
and 328 variables have zero values in their ensemble spread fields, which means lossless compression is required at those points.
\projectname{} can handle these cases natively without the need for special preprocessing.
However, we skip the following two groups of variables.
The first group lacks ensemble spread files, so the pointwise error bounds required by our method are unavailable. 
These fields are typically derived variables that are less important than other variables.
The second group consists of constant boundary-condition fields. 
These fields can be stored separately without needing to be compressed.
Table~\ref{tab:era5_summary} provides a summary of these cases.
The remaining 796 variables have an uncompressed size of 2.35~TB and 
are compressed with \projectname{} under a statistically lossless setup.


\subsubsection{Results}
For all ERA5 variables, our method achieves an overall compression ratio of $37.7\times$.
Across retained variables, single-level variables are compressed by $23.2\times$,
whereas pressure-level variables are compressed by $45.9\times$,
as summarized in Table~\ref{tab:era5_summary}.
Single-level variables are generally less compressible because many of them contain 
finer spatial detail and smaller ensemble spreads.
Pressure-level variables are usually smoother and often have larger admissible pointwise bounds,
which allow higher compression ratios while still satisfying the local uncertainty constraint.
This evaluation shows that the method remains effective in dataset size reduction
when scaled from the most representative variables to 
all ERA5 variables with different characteristics and scientific uses.

\subsection{Robustness to Uncertainty Estimates}\label{sec:uncertainty_robustness}

\begin{figure}[th]
  \centering
  \includegraphics[width=\linewidth]{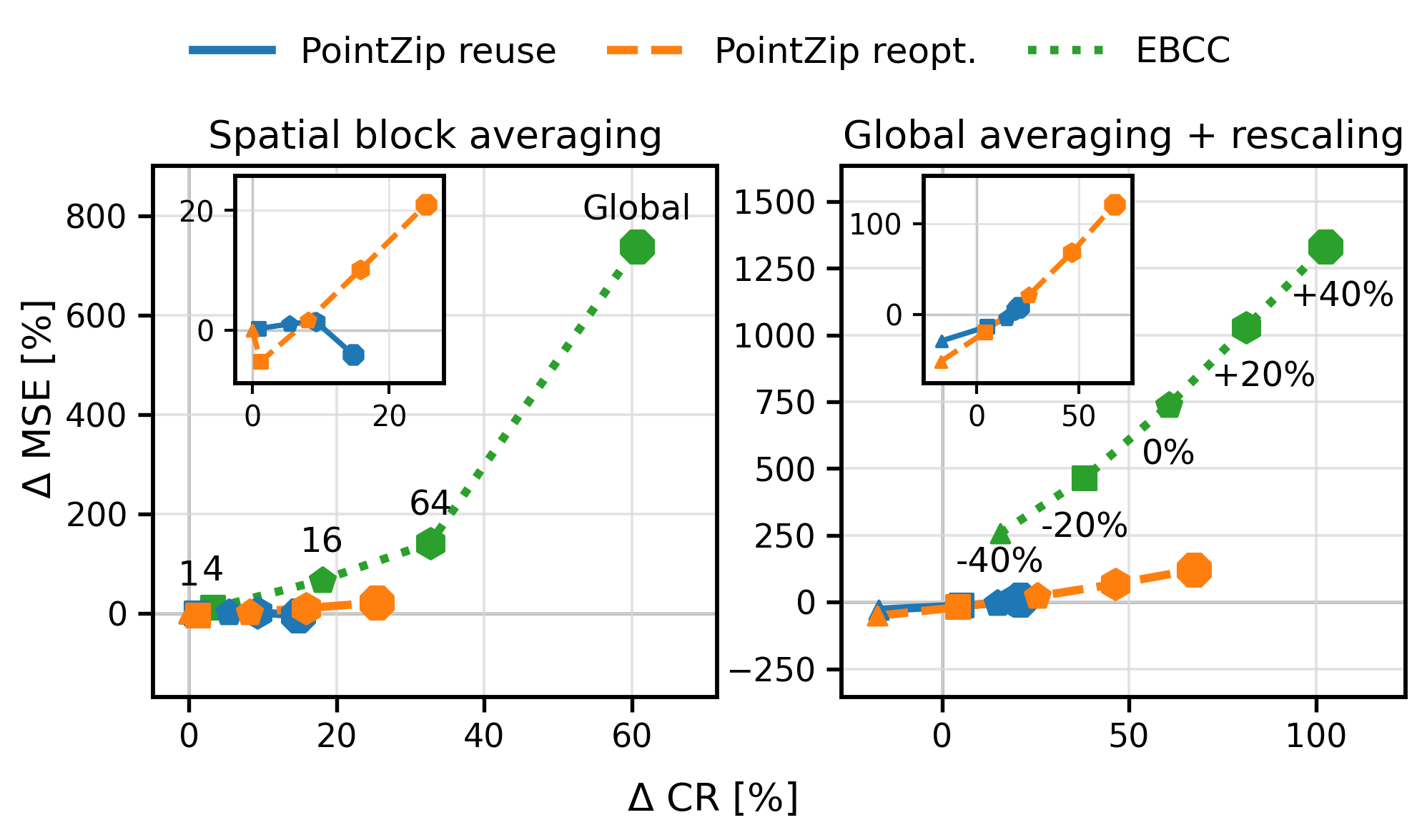}
  \caption{
    Sensitivity to changes in the supplied uncertainty estimates.
    Both axes report percentage changes relative to compression using the original pointwise ERA5 ensemble spread.
    Left: the uncertainty field is progressively smoothed using $4\times4$, $16\times16$, and $64\times64$ spatial block averaging, followed by global averaging.
    Right: the uncertainty field is globally averaged and then rescaled from $-40\%$ to $+40\%$.
    PointZip reuse retains the base-layer compression level selected for the original uncertainty field, whereas PointZip reopt. optimizes this level for each modified field.
  }
  \label{fig:uncertainty_robustness}
\end{figure}

This section answers question (6) by examining how sensitive \projectname{} is to changes in the supplied uncertainty estimates.
Although many climate datasets provide ensemble spread as an uncertainty estimate, this uncertainty estimation can be imperfect.
\projectname{} is designed to maintain a similar compression ratio and reconstruction quality
when the supplied error bounds are changed within a reasonable range.
We use the same ERA5 setup as in Section~\ref{sec:pointwise_common}.
Under changes in the uncertainty estimates, 
we compare \projectname{} with the same base layer compression level,
\projectname{} that reoptimizes the base layer for each modified uncertainty field, and EBCC.

First, we consider the case where the uncertainty field is not accurate for the local variation.
This is a common situation in practice. 
We simulate this by averaging the original ensemble spread by spatial blocks.
This progressively removes local variation while preserving its average magnitude.
This is shown in the left panel of Fig.~\ref{fig:uncertainty_robustness}, 
where we average the uncertainty field over $4\times4$, $16\times16$, and $64\times64$ blocks, followed by global averaging.
\projectname{} shows changes of less than $25\%$ in compression ratio
and less than $21\%$ in reconstruction MSE, even when all spatial
structure in the ensemble spread is removed.
These changes are much smaller than those observed for EBCC,
which shows a $61\%$ change in compression ratio and a change
of more than $700\%$ in reconstruction MSE.
This means that \projectname{} still gives a similar compressed
dataset even when the spatial structure of the uncertainty field is lost.

Second, we consider the case where the dataset provides no uncertainty estimate at all,
so users use a single global uncertainty estimate to compress the data.
This is simulated by replacing the original uncertainty field with its global mean
and scaling it by $-40\%$, $-20\%$, $0\%$, $+20\%$, and $+40\%$,
as shown in the right panel of Fig.~\ref{fig:uncertainty_robustness}.
\projectname{} shows a $67\%$ change in compression ratio
and a change of less than $121\%$ in reconstruction MSE.
For EBCC, the corresponding changes are $103\%$ and more than
$1331\%$, respectively.
This means that for a dataset with no provided uncertainty estimate,
\projectname{} is more robust to the choice of global uncertainty estimate than EBCC.

\section{Related Work}

\paragraph{Lossless Compression for Climate Data}
Common storage backends for climate data include GRIB/GRIB2~\cite{GRIB2}, NetCDF~\cite{NetCDF}, HDF5~\cite{HDF5}, and Zarr~\cite{zarr}.
These systems are typically configured to be lossless or near-lossless.
Besides the DEFLATE family~\cite{zlib,deflate,gzip}, 
scientific storage systems increasingly use codecs such as bzip2~\cite{bzip2}, LZ4~\cite{lz4}, Zstandard~\cite{zstd}, and XZ~\cite{xz} 
to balance throughput and compression ratio.
Some lossless codecs improve the compression ratio of floating-point arrays through transformations.
Blosc applies shuffle-style transformations~\cite{bitshuffle, cblosc}. 
Other floating-point-specific compressors such as FPZIP~\cite{fpzip} and FPC~\cite{fpc} further exploit structure in the floating-point representation. 

\paragraph{Lossy Compression for Climate Data without Error Guarantees}
Classical lossy compression methods optimize rate-distortion performance under average-error metrics such as MSE or PSNR without strict error guarantees. 
This includes image codecs such as JPEG and JPEG2000~\cite{JPEG,JPEG2000}, 
video codecs such as H.264, H.265, and AV1~\cite{ITU_H264,ITU_H265,AV1Spec}, 
and neural compression models~\cite{Balle2017EndToEnd,Minnen2018JointPriors,Cheng2020LearnedCompression,liu2023tcm}.
CRA5~\cite{CRA5} learns a compact latent representation for ERA5 and reconstructs fields with a neural decoder. 
Pellicer-Valero et al.~\cite{VideoCompressionEarthSystemData} use FFmpeg~\cite{ffmpeg} to compress climate data as video sequences 
and report limited degradation in downstream evaluation scores 
when compressed data are used to evaluate neural models trained on the original dataset.

\paragraph{Error-Bounded Lossy Compression for Climate Data}
Following earlier studies~\cite{Woodring2011JPEG2000} on how lossy compression affects downstream scientific analysis,
error-bounded lossy compression has become a major direction in scientific data compression. 
It combines substantial size reduction with explicit reconstruction guarantees~\cite{sz,sz3,zfp,mgard,sperr,ISABELA,di2025survey,ZChecker2017}. 
Representative methods include SZ~\cite{sz}, SZ3~\cite{sz3}, ZFP~\cite{zfp}, MGARD~\cite{mgard}, and SPERR~\cite{sperr}, 
which use prediction-, transform-, or multiresolution-based designs to enforce error control 
while retaining good compression performance. 
These methods typically guarantee a global maximum absolute error bound, and some also support relative error bounds.
Compression methods designed specifically for weather and climate data have also been explored~\cite{ebcc,nature2021,Baker2014}. 
EBCC~\cite{ebcc} is a two-stage compressor that first applies image-based transformations and then compresses residuals in a feedback loop to guarantee error bounds.
Learned compressors, such as SRN-SZ~\cite{srn-sz}, AE-SZ~\cite{ae-sz}, and GraphComp~\cite{GraphComp}, can provide strong rate-distortion performance by exploiting data-specific representations,
but they often require dataset-specific training and an extra non-trainable layer to enforce error bounds.
These algorithms are typically computationally intensive and suffer from limited throughput in real-world applications.

\paragraph{Uncertainty Estimation of ERA5}
All physical measurements are uncertain. 
Modern metrology therefore represents uncertainty probabilistically. 
Under the common Gaussian approximation, a measurement is represented as $\mu \pm \sigma$, 
where $\mu$ is the mean and $\sigma$ is the standard deviation. 
The true value is expected to lie within that interval with approximately $68\%$ probability. 
These conventions are standard in scientific reporting and are reflected in guidance from 
NIST~\cite{Guidelines1994NIST} and JCGM~\cite{JCGM100_2008_GUM, JCGM200_2012_VIM}.

In numerical weather prediction and reanalysis~\cite{4DVar,KalnayBook} 
uncertainty is commonly characterized with ensemble-based methods~\cite{Leutbecher2008EnsembleForecasting}. 
In ECMWF's workflow, uncertainty arises from imperfect initial conditions and uncertainty in the underlying observations, 
and it is propagated through the data-assimilation system that combines observations with the forecast model~\cite{Dee2011ERAInterim}. 
For ERA5, ECMWF provides this through a 10-member ensemble
with perturbed inputs drawn within the uncertainty range of the measurements~\cite{ERA5,ERA5doc}.

\section{Conclusion}

In this work, we introduce \projectname{}, an uncertainty-aware compressor for climate data that uses the ensemble spread as a pointwise error bound. 
This replaces user-chosen global tolerances with a spatially adaptive and scientifically meaningful error target, 
yielding a statistically lossless compressed dataset with respect to data uncertainty.
It enforces pointwise guarantees while remaining practical for standard scientific-data workflows.
Our approach makes global climate simulations at 1km resolution more practical
by reducing storage bandwidth requirements by $72\times$ and enabling an output frequency of 7.5 minutes of simulated time using the same storage bandwidth as an uncompressed output frequency of 3 hours of simulated time.
\projectname{} reaches compression ratios of $69.3\times$ and $37.7\times$ on the most representative variables and all variables in ERA5, respectively.
At matched storage budgets, \projectname{} preserves spectra, extreme events, and air parcel trajectories while introducing less than a $1\%$ increase in RMSE in neural weather forecasts.
These results show that \projectname{} is robust against noise in the uncertainty estimation
and can deliver high-quality compressed data while enabling efficient storage and distribution of large climate datasets.

\section*{Acknowledgment}
This work was supported by the WeatherGenerator project (grant agreement No. 101187947), the ERC PSAP project (grant agreement No. 101002047) and the SERI NCCR Clim+ project. 

\bibliographystyle{IEEEtran}
\bibliography{reference}

\end{document}